\pdfoutput=1

\documentclass[11pt,twoside,a4paper,cmspaper,final,collab]{cms-tdr}
\def\svnVersion{473487P}\def\cmsCernNoTag{CERN-EP-2018-134}\def\cmsCernDate{\today}\def\cmsMessage{Published in Physical Review Letters as \href{http://dx.doi.org/10.1103/PhysRevLett.121.092002}{\doi{10.1103/PhysRevLett.121.092002.}}}

\begin{document}\cmsNoteHeader{BPH-17-008}

\hyphenation{had-ron-i-za-tion}
\hyphenation{cal-or-i-me-ter}
\hyphenation{de-vices}
\RCS$HeadURL: svn+ssh://svn.cern.ch/reps/tdr2/papers/BPH-17-008/trunk/BPH-17-008.tex $
\RCS$Id: BPH-17-008.tex 467341 2018-07-04 08:59:31Z carlosl $

\newlength\cmsFigWidth
\ifthenelse{\boolean{cms@external}}{\setlength\cmsFigWidth{0.95\columnwidth}}{\setlength\cmsFigWidth{0.6\textwidth}}

\newcommand{\Pbgc}{\ensuremath{\chi_{\cPqb}\textrm{(1P)}\xspace}}
\newcommand{\Pbgca}{\ensuremath{\chi_{\cPqb}\textrm{(2P)}\xspace}}
\newcommand{\Pbgcb}{\ensuremath{\chi_{\cPqb}\textrm{(3P)}\xspace}}
\newcommand{\Pbgcib}{\ensuremath{\chi_{{\cPqb}1}\textrm{(3P)}\xspace}}
\newcommand{\Pbgciib}{\ensuremath{\chi_{{\cPqb}2}\textrm{(3P)}\xspace}}
\newcommand{\Pbgcjb}{\ensuremath{\chi_{{\cPqb}J}\textrm{(3P)}\xspace}}
\newcommand{\Pbgcjn}{\ensuremath{\chi_{{\cPqb}J}\xspace}}
\newcommand{\Pcgc}{\ensuremath{\chi_{\cPqc}\xspace}}
\newcommand{\Pcgcin}{\ensuremath{\chi_{{\cPqc}1}\xspace}}
\newcommand{\Pcgciin}{\ensuremath{\chi_{{\cPqc}2}\xspace}}
\newcommand{\Pcgcib}{\ensuremath{\chi_{{\cPqc}1}\textrm{(2P)}\xspace}}
\newcommand{\Pbgcn}{\ensuremath{\chi_{\cPqb}\xspace}}
\newcommand{\Pcgciz}{\ensuremath{\chi_{{\cPqc}1}\xspace}}
\newcommand{\Pbgnc}{\ensuremath{\chi_{\cPqb}\textrm{(1P)}\xspace}}
\newcommand{\Pbgnca}{\ensuremath{\chi_{\cPqb}\textrm{(2P)}\xspace}}
\newcommand{\PXb}{\ensuremath{X_{\cPqb}\xspace}}
\newcommand{\PXBelle}{\ensuremath{X\textrm{(3872)}\xspace}}
\newcommand{\PDDbar}{\ensuremath{\PD\overline{\PD}\xspace}}
\newcommand{\PBBbar}{\ensuremath{\PB\overline{\PB}\xspace}}
\newcommand{\PDDstar}{\ensuremath{\PD\overline{\PD}{}^*\xspace}}
\newcommand{\PBBstar}{\ensuremath{\PB\overline{\PB}{}^*\xspace}}

\cmsNoteHeader{BPH-17-008}
\title{\texorpdfstring
{Observation of the \Pbgcib\ and \Pbgciib\
and measurement of their masses}
{Observation of the chib1(3P) and chib2(3P)
and measurement of their masses}}

\date{\today}

\abstract{The \Pbgcib\ and \Pbgciib\ states are observed through their $\PgUc \, \gamma$ decays, using an event sample of proton-proton collisions collected by the CMS experiment at the CERN LHC. The data were collected at a center-of-mass energy of 13\TeV and correspond to an integrated luminosity of 80.0\fbinv. The \PgUc\ mesons are identified through their dimuon decay channel, while the low-energy photons are detected after converting to \EE pairs in the silicon tracker, leading to a \Pbgcb\ mass resolution of 2.2\MeV. This is the first time that the $J = 1$ and 2 states are well resolved and their masses individually measured: $10\,513.42 \pm 0.41 \stat \pm 0.18 \syst\MeV$ and $10\,524.02 \pm 0.57 \stat \pm 0.18 \syst\MeV$; they are determined with respect to the world-average value of the \PgUc\ mass, which has an uncertainty of 0.5\MeV. The mass splitting is measured to be $10.60 \pm 0.64 \stat \pm 0.17 \syst\MeV$.}

\hypersetup{%
pdfauthor={CMS Collaboration},%
pdftitle={Observation of the chib1(3P) and chib2(3P) and measurement of their masses},%
pdfsubject={CMS},%
pdfkeywords={CMS, physics, quarkonium production, quarkonium spectroscopy}}

\maketitle

Although quantum chromodynamics (QCD) is well established as the theory of the strong interaction,
a complete understanding of the (nonperturbative) processes that lead to the binding of
quarks and gluons into hadrons is still
lacking~\cite{bib:QWG, bib:Brambilla:2010cs,bib:Brambilla:2014jmp}.
The bottomonium family, comprised of beauty quark-antiquark bound states, \bbbar,
plays a special role in understanding how the strong force binds quarks into hadrons
because the large quark mass allows two important theoretical simplifications.
First, the hard-scattering production of a proto-quarkonium quark-antiquark pair can be described
in perturbation theory~\cite{bib:Nayak:2005rw,bib:Nayak:2005rt,bib:Nayak:2006fm}.
Second, the binding of the quark-antiquark pair can be described
in terms of lattice-calculable nonrelativistic
potentials~\cite{bib:Brambilla:1999xf,bib:Brambilla:2000gk,bib:Brambilla:2004jw}.
Particularly stringent tests of current theories of quarkonium production can be achieved by examining
the individual spin states of the quarkonium
multiplets~\cite{bib:Bodwin:1994jh,bib:Bodwin:2015iua,bib:Faccioli:2017hym,bib:Faccioli:2018uik,bib:Faccioli:2018gol}.

The \Pbgcb, observed at a mass of 10.5\GeV by the ATLAS, D0, and LHCb
Collaborations~\cite{bib:ATLAS,bib:D0,bib:LHCb,bib:LHCb_chib3Pmass},
is especially interesting given that its properties could be affected by the
proximity of the open-beauty (\PBBbar) threshold.
Measurements of the masses of the \Pbgcjb\ triplet states,
with total angular momentum $J = 0$, 1, and 2,
probe details of the \bbbar\ interaction and test theoretical treatments
of the influence of open-beauty states on the bottomonium spectrum.
These measurements may also help clarify the nature of several unexpected
charmonium-like states, including the enigmatic \PXBelle~\cite{bib:Choi:2003ue}.
Contending interpretations include the possibility that it is
a mixture of a \Pcgcib\ state and a \PDDstar\ molecule or a compact
tetraquark~\cite{bib:Karliner:2017qhf,bib:Esposito:2016noz,bib:Olsen:2014qna}
or that it is the \Pcgcib, modified by strong-interaction effects
associated with the coincident \PDDstar\ threshold~\cite{bib:Eichten:2005ga}.
The bottomonium analogues of the \Pcgcib\ and \PXBelle\ states
would be the (\bbbar) \Pbgcib\ state
and a possible \PXb\ state at the \PBBstar\ threshold.
Confirming that the \Pbgcib\ is well below the open-beauty threshold
would suggest differences with the charmonium system,
where the \Pcgcib\ state is expected approximately 100\MeV above the \PDDbar\
threshold~\cite{bib:Eichten:2004uh}.
Among various possibilities, the 10.5\GeV peak could be the \PXb\
or a mixture of the \Pbgcib\ and the \PXb~\cite{bib:Karliner:2014lta};
it could also simply be the conventional (unresolved) \Pbgcb,
in which case an hypothetical \PXb\ might exist with a mass close to the \PBBstar\ threshold.
The observation of a doublet structure in the 10.5\GeV peak and a
precise measurement of the mass splitting should confirm the nature of the state
and clarify the existence or absence of effects induced by the nearby open-beauty threshold.

This Letter reports the first observation of resolved \Pbgcib\ and \Pbgciib\ states,
and the measurement of their masses.
The analysis uses the $\PgUc \, \gamma$ decay channel,
with the \PgUc\ decaying to a dimuon and the photon converting into an \EE pair.
It is based on \Pp\Pp\ data samples collected at the CERN LHC by the CMS experiment,
at a center-of-mass energy of 13\TeV, in 2015, 2016, and 2017,
corresponding to integrated luminosities of 2.7, 35.2, and 42.1\fbinv,
respectively~\cite{bib:CMS-PAS-LUM-15-001,bib:CMS-PAS-LUM-17-001,bib:CMS-PAS-LUM-17-004}.
As happens in the \Pcgc, \Pbgc, and \Pbgca\ cases,
the $J = 0$ state of the \Pbgcb\ multiplet is expected to have a negligible radiative-decay branching fraction
and not be observable in the present data sample.

The central feature of the CMS apparatus is a superconducting solenoid of 6\unit{m} internal diameter,
providing a magnetic field of 3.8\unit{T}.
Within the solenoid volume are a silicon pixel and strip tracker, a lead tungstate crystal electromagnetic calorimeter,
and a brass and scintillator hadron calorimeter, each composed of a barrel and two endcap sections.
Forward calorimeters extend the pseudorapidity coverage provided by the barrel and endcap detectors.
Muons are detected in gas-ionization chambers embedded in the steel flux-return yoke outside the solenoid.
A more detailed description of the CMS detector, together with a definition of the coordinate system used and
the relevant kinematic variables, can be found in Ref.~\cite{bib:CMSJINST}.

The data used in this analysis were collected using a two-level trigger system~\cite{bib:CMStrigger}.
The first level consists of custom hardware processors and
uses information from the muon system to select events with two muons.
The high-level trigger requires an opposite-sign muon pair of invariant mass within 8.5--11.5\GeV,
a dimuon vertex fit $\chi^2$ probability larger than 0.5\%,
and a distance of closest approach between the two muons smaller than 0.5\unit{cm}.
The trigger also requires dimuon transverse momentum $\pt > 7.9$\GeV (2015--2016) or 11.9\GeV (2017),
and dimuon rapidity $\abs{y} < 1.25$ (2015--2016) or $\abs{y} < 1.5$ (2017).
The analysis uses photons detected through their conversions to \EE pairs,
following the data reconstruction and selection procedures used in
Refs.~\cite{bib:BPH11010,bib:BPH13005}.

\begin{figure}[ht!]
\centering
\includegraphics[width=0.94\cmsFigWidth]{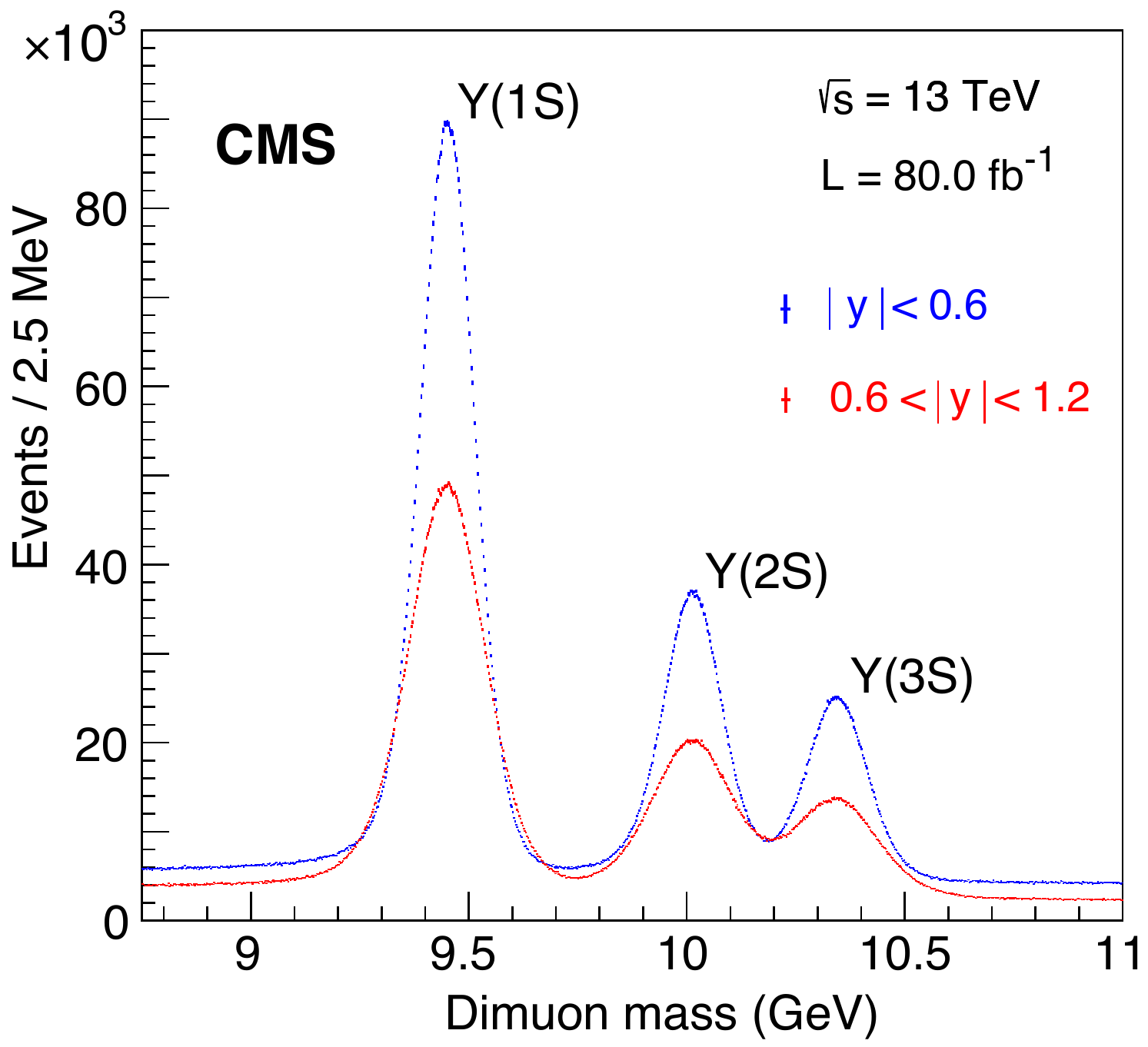}
\caption{The dimuon invariant mass distribution, in two equidistant $\abs{y}$ ranges.
The midrapidity dimuons have a significantly better mass resolution.}
\label{fig:dimuonMass}
\end{figure}

The muon track must have more than five hits in the tracker,
at least one of them being in a pixel detector layer.
The muons selected offline must match, in pseudorapidity and azimuthal angle,
those that triggered the detector readout.
They are combined to form \PgU\ candidates, which are
kept for further processing if $\abs{y} < 1.2$ and $\pt > 14$\GeV.
The selected dimuon sample contains about 10 million \PgUa, 3.9 million \PgUb, and 2.6 million \PgUc.
Figure~\ref{fig:dimuonMass} shows the invariant mass distributions of the selected dimuons,
in two halves of the covered rapidity range.
Fitting such distributions in fine $\abs{y}$ bins reveals that
the dimuon mass resolution $\sigma_m$ varies quadratically
from 60\MeV at $y=0$ to 120\MeV at $\abs{y} = 1.2$.
The background in the mass distribution of the \Pbgcb\ candidates is reduced by
selecting dimuons with invariant mass between
$M(\PgUc) - n_\sigma \, \sigma_m(y)$ and $M(\PgUc) + 2.5 \, \sigma_m(y)$,
where $M(\PgUc)$ is the world-average \PgUc\ mass~\cite{bib:PDG}.
The low-mass edge of the \PgUc\ signal window is defined by
$n_\sigma = 2$ for $\abs{y} < 0.9$ and $n_\sigma = 1.5$ for $0.9 < \abs{y} < 1.2$,
to minimize the contamination from the \PgUb\ resonance.

Photon candidates are formed from two oppositely-charged tracks,
of which one has at least four tracker hits and the other at least three.
The tracks must have a small angular separation,
a small distance of closest approach,
a conversion vertex at least 1.5\cm away from the beam axis,
and a $\chi^2$ probability of the kinematic fit
imposing zero mass and a common vertex that exceeds 0.05\%.
A more detailed account of the selection criteria is given in Ref.~\cite{bib:BPH13005}.
Only photons with pseudorapidity $\abs{\eta} < 1.2$ and
$\pt > 500\MeV$ are kept.

The dimuon is combined with the converted photon to form the \Pbgcb\ candidate.
A kinematic fit of the dimuon-photon system is performed with the following conditions:
the mass of the dimuon is fixed to the \PgUc\ world-average mass, 10.3552\GeV~\cite{bib:PDG};
the electron-positron pair is constrained to have a common vertex and zero mass;
and the two muons and the photon are constrained to have a common vertex.
The \Pbgcb\ candidate is kept if the $\chi^2$ probability of the kinematic fit exceeds 1\%.
Two or more candidates are found in about 1\% of the events; only the one with the best fit is retained.

\begin{figure}[ht]
\centering
\includegraphics[width=\cmsFigWidth]{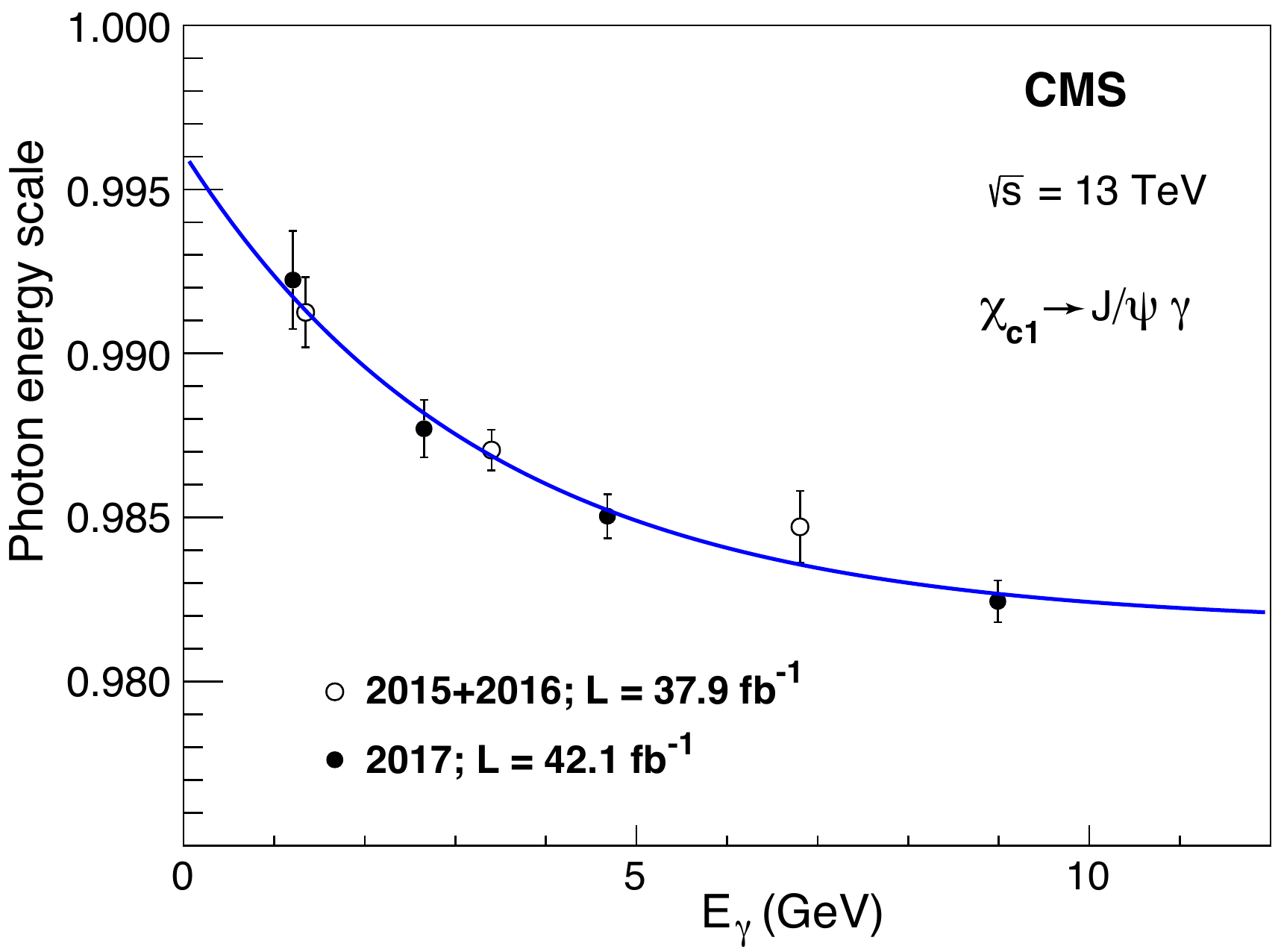}
\caption{The PES as a function of the measured photon energy,
obtained using $\Pcgciz \to \PJGy \, \gamma$ decays
from the 2015--2016 (open circles) and 2017 (filled circles) data samples.
The points are drawn at the average $E_{\gamma}$ in each bin.
The curve represents the parametrization mentioned in the text.}
\label{fig:PES_vs_Egamma}
\end{figure}

To accurately measure the invariant mass of the \Pbgcb\ candidate,
the photon energy scale (PES) must be calibrated.
The PES, defined as the ratio between the reconstructed and true energy,
is measured using a sample of $\Pcgciz \to \PJGy \, \gamma \to \mu^+\mu^- \, \gamma$ events, through the ratio
$[ m_{\mu\mu\gamma}^2 -m_{\mu\mu}^2 ] \,/\, [ M(\Pcgciz)^2 - M(\PJGy)^2 ]$,
where $m_{\mu\mu\gamma}$ and $m_{\mu\mu}$ are the
$\mu\mu\gamma$ and $\mu\mu$ invariant masses,
and $M(\Pcgciz)$ and $M(\PJGy)$ are the world-average masses~\cite{bib:PDG} of the
\Pcgciz\ and \PJGy\ states.
The values are obtained in several bins of photon energy,
profiting from a large $\PJGy \to \mu\mu$ data sample collected
in the same running periods as the $\PgU \to \mu\mu$ data.
The energy spectrum of the $\Pcgciz \to \PJGy \, \gamma$ photons
covers the range relevant for the $\PgU \, \gamma$ analysis.
The PES values, shown in Fig.~\ref{fig:PES_vs_Egamma}
as a function of the measured photon energy, $E_{\gamma}$,
are parametrized with the function $p_0 + p_1 \, \exp(-E_{\gamma}/p_2)$,
where $p_0$, $p_1$, and $p_2$ are free parameters in the fit.
The resulting function is then used for the event-by-event
correction of the photon energy in the computation of the
$\PgU \, \gamma$ invariant mass.

\begin{figure}[ht!]
\centering
\includegraphics[width=\cmsFigWidth]{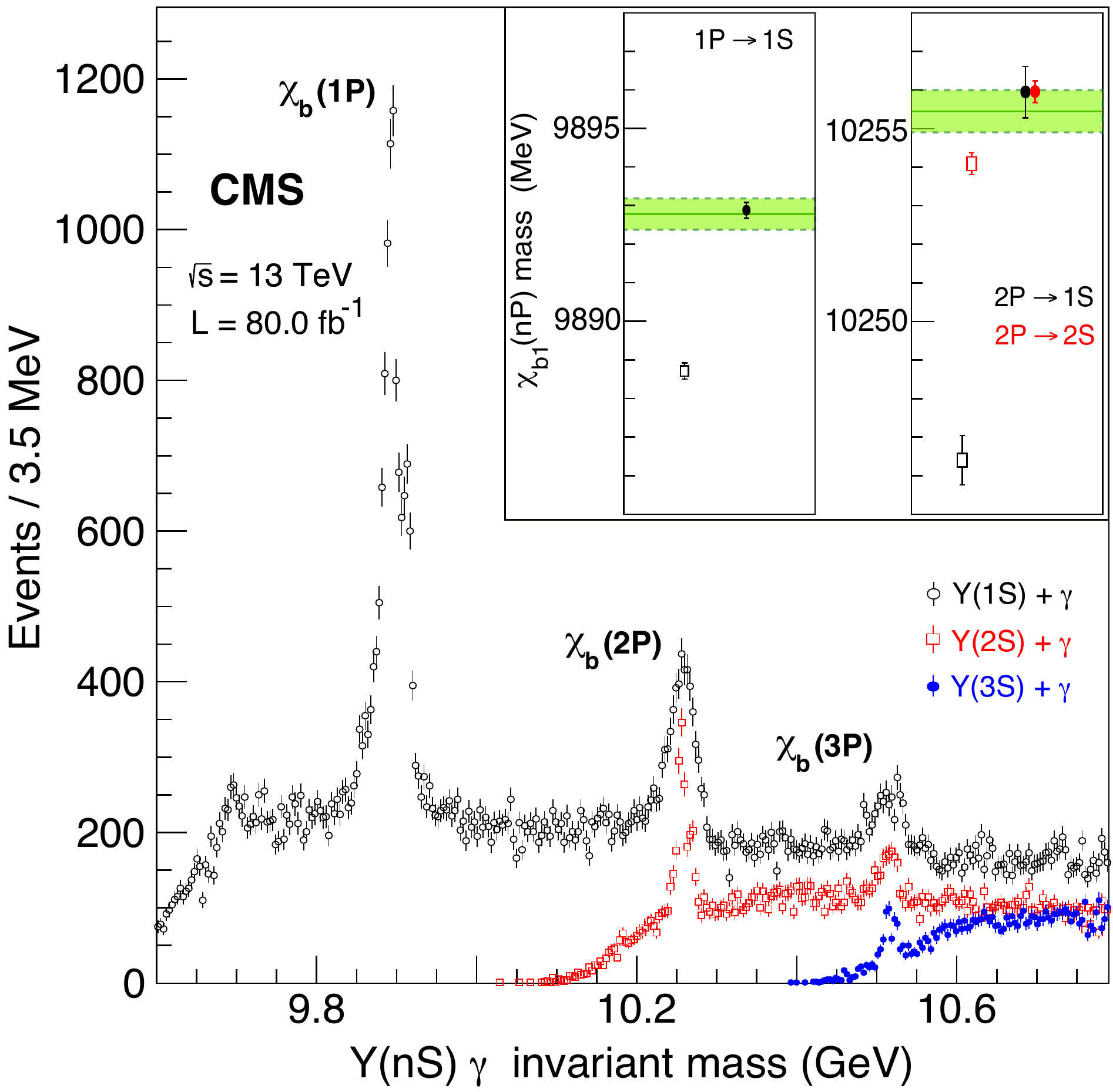}
\caption{The invariant mass distributions of the $\Pbgcjn \to \PgU \textrm{(nS)} \, \gamma$ candidates
($\textrm{n} = 1, 2, 3$), after the PES correction.
The inset shows the \Pbgci\ and \Pbgcia\ masses fitted before (open squares) and after (filled circles) the PES correction,
with vertical bars representing the statistical uncertainties.
The world-average values~\cite{bib:PDG} are shown by the horizontal bands,
with dashed lines representing their total uncertainties.}
\label{fig:chib_mass_values_vs_PDG}
\end{figure}

Figure~\ref{fig:chib_mass_values_vs_PDG} shows the PES-corrected
$\PgU\textrm{(nS)}$-photon invariant mass distributions, with $\textrm{n} = 1, 2, 3$.
The $\PgUa \, \gamma$ and $\PgUb \, \gamma$ events are
selected with the same criteria as used for the $\PgUc \, \gamma$ events,
except that the dimuon invariant mass is required to be between
$M(\PgUa) - 2.5 \, \sigma_m(y)$ and $M(\PgUa) + 2 \, \sigma_m(y)$
and within $M(\PgUb) \pm 2 \, \sigma_m(y)$, respectively.

The prominent \Pbgnc\ and \Pbgnca\ peaks seen in the
$\PgUa \, \gamma$ and $\PgUb \, \gamma$ distributions
in Fig.~\ref{fig:chib_mass_values_vs_PDG} are fit using a
procedure analogous to the one described in the next paragraph.
The resulting \Pbgci\ and \Pbgcia\ masses are in agreement with the world-average
values~\cite{bib:PDG}, as shown in the inset,
confirming the validity of the PES correction function.

\begin{figure}[ht!]
\centering
\includegraphics[width=\cmsFigWidth]{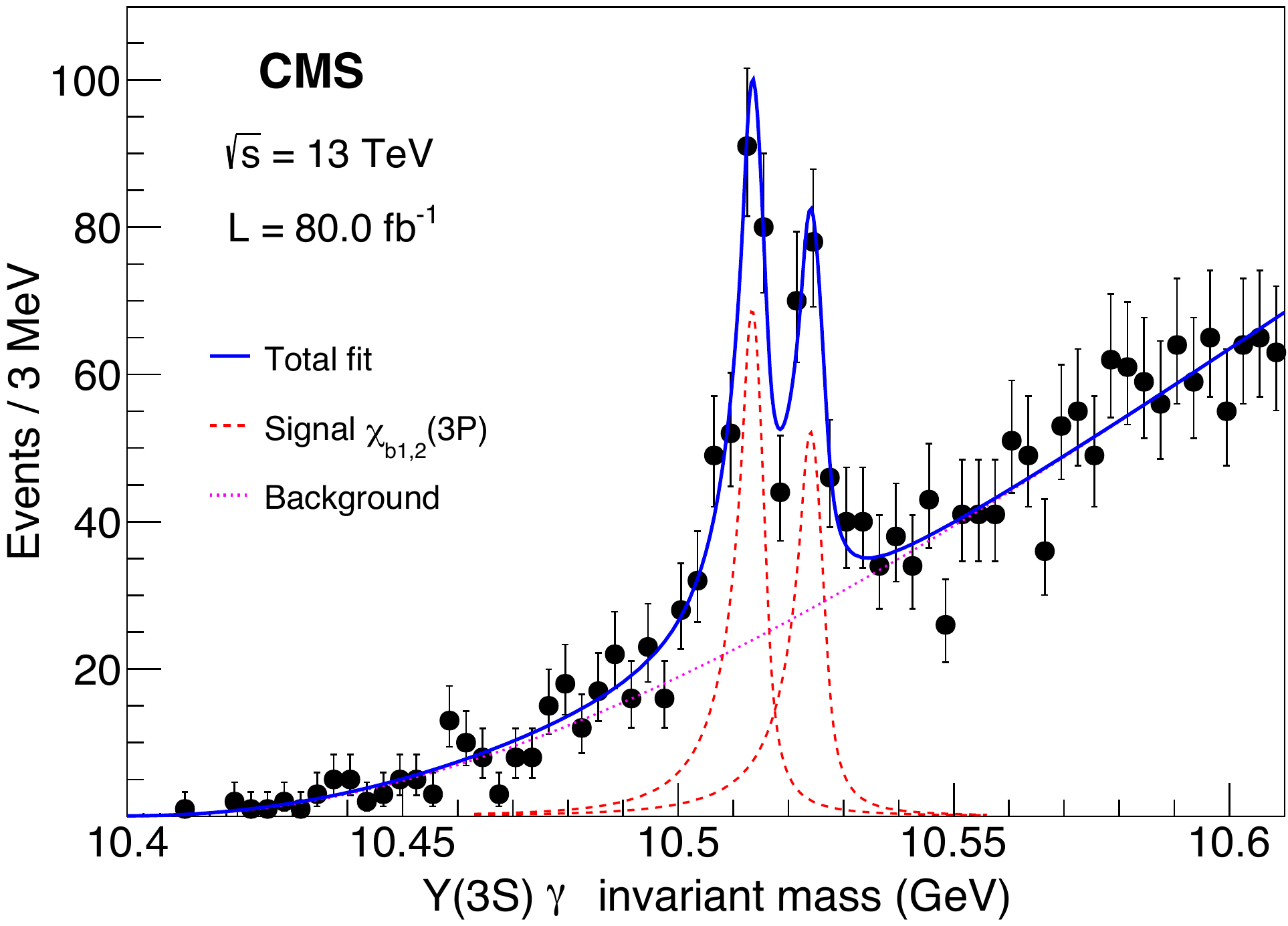}
\caption{The invariant mass distribution of the $\Pbgcjb \to \PgUc \, \gamma$ candidates.
The vertical bars are the statistical uncertainties.
The curves represent the fitted contributions of the two signal peaks, the background, and their sum.}
\label{fig:chibMass}
\end{figure}

Figure~\ref{fig:chibMass} shows the $\PgUc \, \gamma$ invariant mass distribution
along with the result of an unbinned extended maximum-likelihood fit.
The background is described by $(m - q_0)^\lambda \, \exp{[\nu(m - q_0)]}$,
where $m$ is the \Pbgcb\ candidate invariant mass,
$\lambda$ and $\nu$ are free parameters, and $q_0$ is fixed to 10.4\GeV.
The \Pbgcib\ and \Pbgciib\ signal peaks are modeled
with a double-sided Crystal Ball function~\cite{bib:CB},
which complements a Gaussian core with low- and high-mass power-law tails,
defined by the transition points ($\alpha_{\textrm{L,H}}$) and the power-law exponents
($n_{\textrm{L,H}}$).
The tails of the signal functions, identical for both peaks, are defined by the
parameters \mbox{$n_\textrm{L} = 3$} and \mbox{$\alpha_\textrm{L} = 0.6$}, for the low-mass tail,
and by \mbox{$n_\textrm{H} = 2$} and \mbox{$\alpha_\textrm{H} = 1.4$}, for the high-mass tail.
These values reflect studies of simulated distributions, generated with \PYTHIA~8.230~\cite{bib:Pythia},
complemented by \EVTGEN~1.6.0~\cite{bib:EvtGen} to simulate the quarkonium decays
and by \PHOTOS~3.61~\cite{bib:PHOTOS2} for the modeling of final-state radiation.
The generated events undergo a full simulation of the detector response,
according to the implementation of the CMS detector within \GEANTfour~\cite{bib:GEANT};
the samples include multiple \Pp\Pp\ interactions in the same or nearby beam crossings.
The simulation studies show that the resolution of the $\PgU \, \gamma$ mass measurement
is linearly proportional to the difference between the mass of the parent \mbox{P-wave} state
and the mass of the daughter \mbox{S-wave} state, so that one can
impose a linear relationship between the Gaussian widths of the two signal shapes:
$\sigma_2 / \sigma_1 =
[M(\Pbgciib) - M(\PgUc)] \,/\, [M(\Pbgcib) - M(\PgUc)]$.
This relation assumes that the natural widths of the resonances are negligible
with respect to the instrumental resolution.
Fitting without this constraint gives a $\sigma_2 / \sigma_1$ ratio
in agreement with the assumption, albeit with a large uncertainty.

The fitted number of signal events is $372 \pm 36$ and
the fit $\chi^2$ is 46, for 57 degrees of freedom.
The masses of the two resonances are measured to be
$10\,513.42 \pm 0.41$ and $10\,524.02 \pm 0.57\MeV$,
where the uncertainties are statistical only.
The corresponding mass difference is $\Delta M = 10.60 \pm 0.64\MeV$,
where the statistical uncertainty takes into account the correlation 
between the two fitted mass values.
The mass resolution of the low-mass peak is $2.18 \pm 0.32\MeV$,
which agrees with the expectations from simulation studies.
The corresponding resolutions in the $\PgUa \, \gamma$ and $\PgUb \, \gamma$ mass distributions
are 7 and 15\MeV, respectively, justifying why only the $\PgUc \, \gamma$ distribution
is used in this analysis.
The local significance of the double-peak structure was evaluated for several fixed values of $\Delta M$
using a likelihood ratio of two hypotheses,
one of them fixing the yield of the second peak to zero:
it exceeds nine standard deviations in the range $9 < \Delta M < 12\MeV$.

The mass measurements are expected to be essentially insensitive to the event selection criteria.
The analysis was repeated splitting the data sample into subsamples,
using different dimuon rapidity or \pt ranges, or different data collection periods.
The results are also consistent when
the photon $\pt$ thresholds are varied between 400 and 600\MeV,
the dimuon \pt\ thresholds are varied between 12 and 16\GeV,
a broader \PgUc\ mass window is used, $M(\PgUc) \pm 2.5 \, \sigma_m(y)$,
and the minimum dimuon-photon 4-track vertex-fit $\chi^2$ probability is increased to 1.5\%.
Given the absence of significant changes in the results,
the systematic uncertainty related to the selection criteria is considered negligible.
There is also no significant change in the results if the $\sigma_2 / \sigma_1$ ratio
is left free in the fit.

A systematic uncertainty is assigned to account for the fact that the parameters
$\alpha_\textrm{L}$, $\alpha_\textrm{H}$, and $q_0$ are fixed in the
signal and background fit models.
The measured mass distribution was refitted 1000 times, each time with different
values of those parameters, randomly generated according to Gaussian distributions with
nominal mean values and standard deviations reflecting their (correlated) uncertainties.
The $\alpha_\textrm{L}$ and $\alpha_\textrm{H}$ uncertainties are evaluated
as the difference between the fitted values from the measured and simulated \Pbgci\ peaks
in the $\PgUa \, \gamma$ mass distribution,
while the $q_0$ uncertainty is evaluated from a fit to the data
leaving $q_0$ as a free parameter.
The rms of the distribution of the 1000 fit results is taken as
the corresponding uncertainty.
The choice of the analytical function describing the background shape induces
a systematic uncertainty that is evaluated by redoing the fit with two alternative options:
a power-law function, $(m-q_0)^\lambda$ with $q_0$ fixed to 10.4\GeV, and
a Chebychev polynomial of second order.
The total fit-model systematic uncertainty is 0.05\MeV,
both in the mass and mass difference measurements.

The uncertainty in the final results reflecting the precision of the PES correction
function is evaluated with pseudo-experiments, randomly generating 400 correction functions
by drawing new values for its parameters from suitable Gaussian functions,
respecting the corresponding covariance matrix to account for the correlations among the parameters.
The uncertainty associated with the choice of a specific function to fit the
photon energy dependence of the PES is evaluated
by using a constant correction factor,
taken as the average correction in the range ($E_\gamma < 2$\GeV)
relevant for the photons emitted in the $\Pbgcb \to \PgUc \, \gamma$ decays.
The systematic uncertainty reflecting the PES correction is 0.16\MeV
for $\Delta M$ and 0.17\MeV for $M(\Pbgcjb)$.

The total systematic uncertainties are obtained by adding the individual terms in quadrature.
The invariant mass of the \Pbgcn\ candidates is determined by fixing the
dimuon mass to the world-average \PgUc\ mass~\cite{bib:PDG},
presently affected by an uncertainty of 0.5\MeV.
The $\Delta M$ measurement is insensitive to this uncertainty.
The mass difference between the two states is measured to be
$\Delta M = 10.60 \pm 0.64 \stat \pm 0.17 \syst\MeV$,
while the two masses are determined to be
$10\,513.42 \pm 0.41 \stat \pm 0.18 \syst$ and
$10\,524.02 \pm 0.57 \stat \pm 0.18 \syst\MeV$.

These values can be compared to the predictions of theoretical calculations~\cite{bib:Li:2009nr,
bib:Dib:2012vw, bib:Liu:2011yp, bib:Bhaghyesh:2013dga, bib:Wei-Zhao:2013sta,
bib:Repko:2012rk, bib:Ferretti:2014xqa, bib:Ferretti:2013vua, bib:Godfrey:2015dia,
bib:Segovia:2016xqb, bib:Lu:2016mbb, bib:Deng:2016ktl}.
Out of 19 $\Delta M$ predictions, 18 range from 8 to 18\MeV,
mostly depending on the potentials describing the \bbbar\ nonperturbative interaction.
The only exception gives $M(\Pbgciib) - M(\Pbgcib) = -2\MeV$,
the negative sign reflecting the coupling with the open-beauty threshold,
whose proximity could have a striking influence on the \Pbgcjb\
splitting~\cite{bib:Ferretti:2014xqa, bib:Ferretti:2013vua}.
The measurement reported in this Letter shows that the mass gap between the
$J = 1$ and 2 states is significantly larger than 2\MeV, an observation that strongly disfavours the
breaking of the conventional pattern of splittings as presented in that specific calculation
and supports the standard mass hierarchy, where the $J=2$ state is heavier than the $J=1$ state.
It is also worth noting that the measured $\Delta M$ agrees with the value of 10.5\MeV that was
assumed in Ref.~\cite{bib:LHCb_chib3Pmass}.

In summary, data samples of \Pp\Pp\ collisions at $\sqrt{s} = 13\TeV$,
collected by CMS in the years 2015--2017,
corresponding to an integrated luminosity of 80.0\fbinv,
were used to measure the invariant mass distribution of
the $\Pbgcb \to \PgUc \, \gamma$ candidates,
with the \PgUc\ mesons detected in the dimuon decay channel and
the photons reconstructed through conversions to \EE pairs.
The measured distribution is well reproduced by the superposition of the
\Pbgcib\ and \Pbgciib\ quarkonium states,
overlaid on a smooth continuum.
This is the first time that the two states are individually observed.
Their mass difference is
$\Delta M = 10.60 \pm 0.64 \stat \pm 0.17 \syst\MeV$,
and their masses, assuming that the $J = 1$ state is the lighter one,
are $M(\Pbgcib) = 10\,513.42 \pm 0.41 \stat \pm 0.18 \syst$
and $M(\Pbgciib) = 10\,524.02 \pm 0.57 \stat \pm 0.18 \syst \MeV$,
having an additional 0.5\MeV uncertainty reflecting the present precision of the world-average \PgUc\ mass.
This measurement fills a gap in the spin-dependent bottomonium spectrum below the open-beauty threshold
and should significantly contribute to an improved understanding of the nonperturbative spin-orbit interactions
affecting quarkonium spectroscopy.

\begin{acknowledgments}

We thank Geoff Bodwin, Estia Eichten, and Chris Quigg for important theoretical input on short notice.
We congratulate our colleagues in the CERN accelerator departments for the excellent performance of the LHC and thank the technical and administrative staffs at CERN and at other CMS institutes for their contributions to the success of the CMS effort. In addition, we gratefully acknowledge the computing centers and personnel of the Worldwide LHC Computing Grid for delivering so effectively the computing infrastructure essential to our analyses. Finally, we acknowledge the enduring support for the construction and operation of the LHC and the CMS detector provided by the following funding agencies: BMWFW and FWF (Austria); FNRS and FWO (Belgium); CNPq, CAPES, FAPERJ, and FAPESP (Brazil); MES (Bulgaria); CERN; CAS, MoST, and NSFC (China); COLCIENCIAS (Colombia); MSES and CSF (Croatia); RPF (Cyprus); SENESCYT (Ecuador); MoER, ERC IUT, and ERDF (Estonia); Academy of Finland, MEC, and HIP (Finland); CEA and CNRS/IN2P3 (France); BMBF, DFG, and HGF (Germany); GSRT (Greece); NKFIA (Hungary); DAE and DST (India); IPM (Iran); SFI (Ireland); INFN (Italy); MSIP and NRF (Republic of Korea); LAS (Lithuania); MOE and UM (Malaysia); BUAP, CINVESTAV, CONACYT, LNS, SEP, and UASLP-FAI (Mexico); MBIE (New Zealand); PAEC (Pakistan); MSHE and NSC (Poland); FCT (Portugal); JINR (Dubna); MON, RosAtom, RAS and RFBR (Russia); MESTD (Serbia); SEIDI, CPAN, PCTI and FEDER (Spain); Swiss Funding Agencies (Switzerland); MST (Taipei); ThEPCenter, IPST, STAR, and NSTDA (Thailand); TUBITAK and TAEK (Turkey); NASU and SFFR (Ukraine); STFC (United Kingdom); DOE and NSF (USA).

\end{acknowledgments}

\bibliography{auto_generated}

\cleardoublepage \appendix\section{The CMS Collaboration \label{app:collab}}\begin{sloppypar}\hyphenpenalty=5000\widowpenalty=500\clubpenalty=5000\vskip\cmsinstskip
\textbf{Yerevan Physics Institute, Yerevan, Armenia}\\*[0pt]
A.M.~Sirunyan, A.~Tumasyan
\vskip\cmsinstskip
\textbf{Institut f\"{u}r Hochenergiephysik, Wien, Austria}\\*[0pt]
W.~Adam, F.~Ambrogi, E.~Asilar, T.~Bergauer, J.~Brandstetter, M.~Dragicevic, J.~Er\"{o}, A.~Escalante~Del~Valle, M.~Flechl, R.~Fr\"{u}hwirth\cmsAuthorMark{1}, V.M.~Ghete, J.~Hrubec, M.~Jeitler\cmsAuthorMark{1}, N.~Krammer, I.~Kr\"{a}tschmer, D.~Liko, T.~Madlener, I.~Mikulec, N.~Rad, H.~Rohringer, J.~Schieck\cmsAuthorMark{1}, R.~Sch\"{o}fbeck, M.~Spanring, D.~Spitzbart, A.~Taurok, W.~Waltenberger, J.~Wittmann, C.-E.~Wulz\cmsAuthorMark{1}, M.~Zarucki
% ...
\vskip\cmsinstskip
% ...
1:  Also at Vienna University of Technology, Vienna, Austria\\
% ...
\end{sloppypar}
\end{document}